\documentclass[final]{aa}
\usepackage{graphics,epsf,epsfig,graphicx}
\usepackage{natbib}
\usepackage{url}

\bibpunct{(}{)}{;}{a}{}{,} % to follow the A&A style

\newcommand{\be}{\begin{equation}}
\newcommand{\ee}{\end{equation}}
\newcommand{\dd}{\mathrm{d}}

\begin{document}

\title{Monte Carlo radiative transfer in protoplanetary disks}

\author{Christophe Pinte\inst{1}, Fran\c{c}ois M\'enard\inst{1}, 
Gaspard Duch\^ene\inst{1}, Pierre Bastien\inst{2}}

\offprints{C. Pinte \\ \email{christophe.pinte@obs.ujf-grenoble.fr}}
\institute{Laboratoire d'Astrophysique de Grenoble, CNRS/UJF UMR~5571, 
414 rue de la Piscine, B.P. 53, F-38041 Grenoble Cedex 9, France
\and 
D\'epartement de physique et Observatoire du Mont-M\'egantic, Universit\'e de Montr\'eal, C. P. 6128, Succ. Centre-ville, Montr\'eal,  QC H3C 3J7, Canada}

\authorrunning{Pinte et al.}
\titlerunning{Monte Carlo radiative transfer in protoplanetary disks}

\date{Received ... / Accepted ...}

\abstract{}{We present a new continuum 3D radiative transfer code,
MCFOST, based on a Monte-Carlo method.  MCFOST can be used to
calculate (i) monochromatic images in scattered light and/or thermal
emission, (ii) polarisation maps, (iii) interferometric visibilities,
(iv) spectral energy distributions and (v) dust temperature distributions
of protoplanetary disks.}{Several improvements to the standard Monte
Carlo method are implemented in MCFOST to increase efficiency and
reduce convergence time, including wavelength distribution
adjustments, mean intensity calculations and an adaptive sampling of
the radiation field. The reliability and efficiency of the code are
tested against 
a previously defined benchmark, using a 2D disk configuration.  No significant
difference (no more than 10\%, and generally much less) is found
between the temperatures and SEDs calculated by MCFOST and by other
codes included in the benchmark.}{  A study of the lowest disk mass
  detectable by \textit{Spitzer}, around young stars, is presented and the colours
of ``representative'' parametric disks are compared to recent IRAC and MIPS \textit{Spitzer} colours of solar-like young
stars located in nearby star forming regions.}{}

\maketitle

%======================================================================
\section{Introduction}

Signatures for the presence of dust are found nearly everywhere in
astrophysics. In the context of star and planet formation, dust is
abundant in molecular clouds and in the circumstellar environments of
a large fraction of stellar objects in the early stages of their
evolution. In the circumstellar disks encircling these young stars,
where planets are thought to form, the interplay between dust and gas
is of paramount importance.

At short wavelengths, dust grains efficiently absorb, scatter, and
polarise the starlight. How much radiation is scattered and absorbed
is a function of both the geometry of the disk and the properties of
the dust. In turns, the amount of absorbed radiation sets the temperature of the
dust (and gas) and defines the amount of radiation that is re-emitted
at longer, thermal, wavelengths.

The last decade has witnessed the improvement of imaging capabilities
with the advent of potent instruments in the optical and
near-infrared, and large millimeter interferometers, providing
detailed views of the disks around young stars. The sensitivity and
wavelength range covered by new instruments is increasing steadily and
the mid- and far-infrared ranges are now being explored efficiently by
the {\sl Spitzer Space Telescope}, and new facilities like Herschel
and ALMA will soon complete the coverage.

With this unprecedented wealth of data, from optical to radio, fine
studies of the dust content and evolution of disks become
possible and powerful radiative transfer (RT) codes are needed to fully
exploit the data.  In this paper we describe such a code, MCFOST.

 That code was used
extensively to produce synthetic images of the scattered light from
disks around young stars. Examples include the circumbinary ring of
GG~Tau \citep{McCabe02, Duchene04}, the large silhouette disk
associated with IRAS 04158+2805 \citep{Menard05}, and an analysis of
the circular polarisation in GSS~30 \citep{Chrysostomo97}. 

In \S2 below, we briefly describe MCFOST. In section \S3, tests and
validation of MCFOST are presented. Two examples of applications are
presented in \S4. Firstly, a study is presented of the minimum
mass of disks detectable by {\sl Spitzer} around young solar-like stars, young
low-mass stars, and young brown dwarfs. Secondly, the colours of
parametric disks are compared to the {\sl Spitzer}
colours, for both IRAC and MIPS, of samples of T Tauri stars located in
nearby star forming regions.

%======================================================================
\section{Description of the numerical code}
MCFOST is a 3D continuum radiative transfer code based on the
Monte-Carlo method. It was originally developed by \cite{Menard} to
model the scattered light in dense dusty media (including linear and
circular polarisations). In this paper, we present an extended version
of the original code that includes dust heating and continuum thermal
re-emission.

\subsection{Geometry of the computation grid}

MCFOST uses a spatial grid that is defined in cylindrical coordinates,
well adapted to the geometry of circumstellar disks. $N_r$
logarithmically-spaced radial grid cells and $N_z$ linearly-spaced
vertical grid cells are used. The size of the vertical cells follows
the flaring of the disk, with a cutoff at 10~times the scale height
for each radius. MCFOST allows for the density distribution to be
defined arbitrarily, in 3 dimensions, with the limitation that within
each cell, quantities are held constant.
The density distribution of the media under consideration can be
described either by using a parametric model (as in
\S~\ref{sec:bench}) or by including any ad hoc density table, calculated 
by other means. In the following, the term ``local'' refers to the
properties of a single cell. 

\subsection{Position-dependent dust distributions}
An explicit spatial dependence of the size (and/or composition)
distribution $f(a,\vec{r})$ is implemented in MCFOST.  The dust
properties are therefore defined locally, i.e., each cell of the disk
may contain its own independent dust population. This allows for
example to model dust settling towards the disk midplane, variations
of the chemical composition of the dust, and/or an increase of ice
mantles from the inner, hot regions to the outer, cold edge of the
disk.

\subsection{Optical dust properties}
The dust optical properties are currently computed with the Mie
theory, i.e., grains are spherical and homogeneous.  The optical
properties at a given position in the disk are derived in accordance
with the local size and composition distributions $f(a,\vec{r})$. The
extinction and scattering opacities are given by \be
\kappa^\mathrm{ext/sca}(\lambda,\vec{r}) =
\int_{a_\mathrm{min}}^{a_\mathrm{max}} \pi a^2
Q_\mathrm{ext/sca}(\lambda,a) f(a,\vec{r}) \,\mathrm{d}a \ee where $
Q_\mathrm{sca}(\lambda,a)$ and $ Q_\mathrm{ext}(\lambda,a)$ are
respectively the scattering and extinction cross sections of a grain
of size $a$ at a wavelength $\lambda$.

\subsection{The radiative transfer scheme in MCFOST}

The Monte Carlo method allows to follow individual monochromatic
``photon packets'' that propagate through the circumstellar
environment until they exit the computation grid.  The propagation
process is governed by scattering, absorption and re-emission events
that are controlled by the optical properties of the medium (opacity,
albedo, scattering phase function, etc...) and by the temperature
distribution. Upon leaving the circumstellar environment, ``photon
packets'' are used to build an SED and/or synthetic images, one image
per wavelength.

\subsubsection{The emission of ``photon packets''}

In MCFOST, two general radiation sources are considered~: photospheric
emission and circumstellar dust thermal emission.

Any number of stars can be considered, at any position in the
calculation grid. The photospheric emission can be represented by (i)
a sphere radiating uniformly or (ii) a limb-darkened sphere or (iii) a
point-like source, whichever is relevant for the problem under
consideration. The spectrum of photospheric photons can follow a
blackbody or any observed or calculated spectrum. Hot and/or cold
spots can be added to the photospheres.

For the dust thermal emission, the density, temperature and opacities
are considered constant within each grid cell. Thermal photons
emission locations are uniformly distributed within each cell and the
emission is assumed to be isotropic.

\subsubsection{Distance between interactions}

It is natural within a Monte Carlo scheme to estimate the distance a
photon packet ``travels'' between two interactions by means of optical
depth (directly related to the density) rather than by physical,
linear distance. From a site of interaction, i.e., scattering and
emission, the optical depth $\tau$ to the next site is randomly chosen
following the probability distribution $p(\tau) = \exp(-\tau)$.

The distance $l$ is computed by integrating the infinitesimal
optical depth $\kappa^\mathrm{ext}(\lambda,\vec{r}) \rho(\vec{r})
\,\mathrm{d}s$ until the following equality is verified~:
\be
\tau_\lambda = \int_0^l \kappa^\mathrm{ext}(\lambda,\vec{r}) \rho(\vec{r}) \,\mathrm{d}s~.
\ee

Once the position of interaction $\vec{r}$ is determined, the
probability that the interaction is a scattering event rather than an 
absorption event is estimated with the local albedo~:
\be
{p_\mathrm{sca} = \mathcal{A}(\lambda,\vec{r}) = \frac{ \int_{a_\mathrm{min}}^{a_\mathrm{max}} \pi a^2
Q_\mathrm{sca}(\lambda,a) f(a,\vec{r}) \,\mathrm{d}a}{\int_{a_\mathrm{min}}^{a_\mathrm{max}} \pi a^2
Q_\mathrm{ext}(\lambda,a) f(a,\vec{r}) \,\mathrm{d}a}}~.
\ee

\subsubsection{Scattering}

MCFOST includes a complete treatment of linear and circular
polarisations by using the Stokes formalism. The state of a light
packet is described by its Stokes vector, with $I$ representing the intensity, $Q$
and $U$ the linear polarisation, and $V$ the circular
polarisation.  The interaction of a photon with a dust particle is
described by a $4\times4$ matrix, the Mueller matrix, ${\bf S}$, also
known as the scattering matrix. Since the matrix is defined
  with respect to the scattering plane, appropriate rotation matrices
  are required to transform the Stokes parameters between multiple
  scattering events. In the simplifying case of Mie
scattering
, the matrix becomes block-diagonal with only 4 different
non-zero elements, see Eq.~\ref{eq:mueller}.

\be
\label{eq:mueller}
\left( \begin{array}{r}
I \\
Q \\
U \\
V \\
\end{array} \right)_{i+1}
=
\left( \begin{array}{rrrr}
S_{11} & S_{12} & 0 & 0 \\
S_{12} & S_{11} & 0 & 0 \\
0    &  0  & S_{33} & S_{34} \\
0 & 0 & -S_{34} & S_{33} \\
\end{array} \right)
\left( \begin{array}{r}
I \\
Q \\
U \\
V \\
\end{array} \right)_{i}
\ee

The direction of scattering is defined by two angles, the scattering
angle $\theta$ and the azimuth angle $\phi$ in spherical
coordinates. Each individual element $S_{ij}$ of the matrix is a
function of the scattering angle and of the wavelength, $S_{ij} =
S_{ij}(\theta,\lambda)$.  The scattering angle $\theta$ is randomly
chosen from the pretabulated cumulative distribution function~: \be
\label{eq:phase_theta}
F(\theta) = \frac{\int_0^\theta S_{11} (\theta')\,\sin \theta' \mathrm{d}\theta'}{\int_0^\pi S_{11} (\theta')\,\sin \theta' \mathrm{d}\theta'}.
\ee 

For linearly unpolarised incoming light packets, i.e., $Q,U = 0$, the
distribution of azimuthal angle $\phi$ is isotropic. For light packets
with a non-zero linear polarisation, $P =\sqrt{Q^2+U^2}/I$, the
azimuthal angle is defined relative to its direction of polarisation
and determined by means of the following cumulative distribution function~:
\be
\label{eq:phase_phi}
F_\theta(\phi) =
\frac{1}{2\pi}\left(\phi-\frac{S_{11}(\theta)-S_{12}(\theta)}{S_{11}(\theta)+S_{12}(\theta)}\,P\,\frac{\sin(2\phi)}{2}\right)
\ee 
where $\theta$ was previously chosen from Eq.~\ref{eq:phase_theta}
\citep{Solc89}.

The Mueller matrix used in Eqs. \ref{eq:mueller},
\ref{eq:phase_theta}, and  \ref{eq:phase_phi} can be defined in two different 
ways : one Mueller matrix per grain size or one mean Mueller matrix per
grid cell. In the latter case, the local Mueller matrix
$S(\lambda,\vec{r})$ is defined by~: 
\be
\label{eq:def_S}
{\bf S}(\lambda,\vec{r}) = \int_{a_\mathrm{min}}^{a_\mathrm{max}}
{\bf S}(\lambda,a) f(a,\vec{r}) \,\mathrm{d}a
\ee
where $S(\lambda,a)$ is the Mueller matrix of a grain of size $a$ at a wavelength $\lambda$.

In the former case, the grain size must be
chosen explicitly for each event following the probability law~:
\be
p(a)\dd a = \frac{\pi a^2
  Q_{\mathrm{sca}}(a)f(a,\vec{r})\,\dd a}{\int_{a_\mathrm{min}}^{a_\mathrm{max}}\pi a^2 Q_{\mathrm{sca}}(a)f(a,\vec{r})\,\dd a}~.
\ee
The two methods are strictly equivalent. They can be used alternatively
to minimize either the memory space or the computation time required.

\subsubsection{Absorption and radiative equilibrium\label{sec:eq_rad}}
The temperature of the dust particles is determined by assuming
radiative equilibrium in the whole model volume, and by assuming that
the dust opacities do not depend on temperature.

The dust thermal balance should take into account the
thermal coupling between gas and dust. In high density regions, e.g.,
close to the disk midplane, the coupling is very good and the gas
temperature is expected to be close to the dust temperature. At the
surface layers of the disk, on the other hand, densities become much
lower and gas-dust thermal exchanges are reduced.

Two extreme assumptions are implemented in MCFOST~: either the
gas-dust thermal exchange is perfect and gas and dust are in local
thermodynamic equilibrium (LTE) throughout the disk, or there is no
thermal coupling between gas and dust at all. In the latter case, each
grain size has its own, different temperature, despite being in
equilibrium with the same radiation field.

Under the assumption of LTE, all dust grains have the same
temperature, which is also equal to the gas temperature. In this case, the
(unique) temperature of the cell is given by the radiative
equilibrium equation~:
\begin{equation}
4\pi\int_0^\infty \kappa_i^\mathrm{abs}(\lambda) B_\lambda(T_i)
\,\dd\lambda = \Gamma_i^\mathrm{abs}
\end{equation}
where $\Gamma_i^\mathrm{abs}$ is the  energy absorption rate. If only
passive heating is considered, this rate
can be written from the mean intensity, leading to~:
\begin{equation}
\label{eq:radequil}
4\pi\int_0^\infty \kappa_i^\mathrm{abs}(\lambda) B_\lambda(T_i)
\,\dd\lambda = 4\pi\int_0^\infty \kappa_i^\mathrm{abs}(\lambda)
J_\lambda \,\dd\lambda.
\end{equation}
Any extra internal source of energy (viscous heating for instance) can
be easily included by adding the corresponding term on the
right hand side of Eq.~\ref{eq:radequil}.

Each time a packet $\gamma$ of a given wavelength $\lambda$ travels
through a cell, the quantity $\Delta l_\gamma$, the distance
travelled by the packet in the cell, is computed. The mean intensity
$J_\lambda$ in the cell is then derived following \citet{Lucy99}:
\be
J_\lambda = \frac{1}{4\pi V_i} \sum_{\gamma} \varepsilon\, \Delta
l_\gamma = \frac{1}{4\pi V_i} \sum_{\gamma} \frac{L_*}{N_\gamma}\,
\Delta l_\gamma
\ee
where $V_i$ is the volume of cell $i$ and $\varepsilon$ the individual
luminosity of a packet. The radiative equilibrium equation can then be
written as~:
\begin{equation} 
\int_0^\infty \kappa_i^\mathrm{abs}(\lambda)
B_\lambda(T_i) \, \dd\lambda = \frac{L_*}{4\pi V_i N_\gamma} \sum_{\lambda,
\gamma} \kappa_i^\mathrm{abs}(\lambda) \Delta l_\gamma~. 
\label{eq:Lucy}
\end{equation}
This method is known as the {\sl mean intensity} method.

When gas and
dust, as well as dust grains of different sizes, are not thermally
coupled,
 there is a different temperature for each grain size and the
radiative equilibrium must be written independently for
each of them~:
\begin{equation}
4\pi\int_0^\infty \kappa_{i}^\mathrm{abs}(\lambda,a) B_\lambda(T_{i}(a))
\,\dd\lambda = \Gamma_i^\mathrm{abs}(a)
\end{equation}
where $\kappa_{i}^\mathrm{abs}(\lambda,a)$, $T_{i}(a)$ and $
\Gamma_i^\mathrm{abs}(a)$ are the
opacity, temperature and energy absorption rate of the dust grains of size $a$ in cell $i$.

The calculation of $\int_0^\infty \kappa_i^\mathrm{abs}(\lambda)
B_\lambda(T_i) \, \dd\lambda = \sigma T^4 \kappa_P / \pi$ ($\kappa_P$
is the Planck mean opacity) is very time consuming and we pretabulate
these values at $N_T = 1\,000$ logarithmically spaced temperatures
ranging from $1$K to the sublimation temperature, which is of order
$1\,500$ K. The temperature $T_i$ of each cell is obtained by interpolation
between the pretabulated temperatures.

To speed up calculations, \citet{Bjorkman01} introduced the concepts
of {\sl immediate reemission} and {\sl temperature correction}. They
are implemented in MCFOST.  In essence, when a packet is absorbed, it
is immediately re-emitted and its wavelength chosen by taking into
account the temperature correction given by the following probability
distribution~:
\be p_\lambda \, \dd\lambda \propto
\kappa_\mathrm{i}^\mathrm{abs}(\lambda)
\left(\frac{\mathrm{d}B_\lambda}{\mathrm{d}T}\right)_{T_i} \, \dd\lambda~.
\label{eq:correct_PDF}
\ee

\citet{Baes05} suggested that this wavelength distribution adjustment 
method may not be entirely safe when used in combination with the
{\sl mean intensity} writing of the absorption rate described above.
Instead, MCFOST determines the cell temperatures $T_i$ used in
Eq. \ref{eq:correct_PDF} by solving~:
\begin{equation} 
\int_0^\infty \kappa_i^\mathrm{abs}(\lambda)
B_\lambda(T_i) \, \dd\lambda = \frac{L_*}{4\pi V_i N_\gamma} N_{\gamma\
  \mathrm{abs}, i}
\label{eq:Lucy_alt}
\end{equation}
where $N_{\gamma\ \mathrm{abs}}$ is the current number of
absorbed photons in cell $i$. Eq. \ref{eq:Lucy} is used only at the
end,
 when all photons have been propagated, to improve the accuracy of
the temperature determination.  Using Eq.~\ref{eq:Lucy_alt} instead of
Eq.~\ref{eq:Lucy} when computing temperature for
Eq.~\ref{eq:correct_PDF} is in no way a limitation because, at this
stage of the computation, the aim is to compute the radiation field
and not the whole temperature structure. Of course, the temperature in
optically thin cells will not be precisely determined because little
absorption occurs in these cells. On the other hand, as a consequence,
little reemission will also occur and the contribution of these
reemission events will remain very small. By construction of the
\cite{Bjorkman01} method, temperature is best converged in the cells
that contribute most to the radiation field. So, when all photons have
been propagated, temperature corrections are no longer needed and the
temperature of all cells can be computed following the \cite{Lucy99}
method, providing a much higher final accuracy.

\subsubsection{SED calculation: sampling of the radiation field\label{sec:sampling}}

To improve its computational efficiency when producing SEDs, MCFOST
uses different samplings of the radiation field, i.e., different
strategies to set the energy of a photon packet.  The energy of a
packet, hence the number of photons it contains, is chosen and
optimised depending on the goals of the calculation. On the one hand,
the convergence of the temperature distribution is optimized when all
photon packets have the same {\sl energy}, independently of their
wavelengths. This procedure insures that more photon packets are
emitted in the more luminous bins. On the other hand, the computation
of SEDs itself is more efficient when it is the {\sl number} of photon
packets that is held constant for all wavelength bins instead.  In
that case it is the energy of the packets that is wavelength
dependent.  Thus, MCFOST is made more efficient by computing the
temperature and SEDs with a two-step process:
\begin{itemize}
\item {\bf Step 1} is the temperature determination. Photon packets
are generated and calculated one by one at the
stellar photosphere and propagated until they exit the computation
grid, i.e., until they exit the circumstellar medium.  Upon scattering,
the propagation vector of a packet is modified, but not its
wavelength. Upon absorption however, packets are immediately
re-emitted, in situ and isotropically, but at a different wavelength
calculated according to the temperature of the grid cell. For
this re-emission process, the concept of {\sl immediate reemission}
and the associated {\sl temperature correction method} proposed by
\cite{Bjorkman01} are used. In this step, all photon packets have the
same luminosity
  \begin{equation}
    \label{eq:E_step1}
    \epsilon = L_*/N_{\gamma step1} 
  \end{equation}
where $L_*$ is the bolometric luminosity of the star and $ N_{\gamma
step1}$ the number of packets generated,
 and are randomly scattered/absorbed within the disk. The
concept of {\sl mean intensity} suggested by \cite{Lucy99} is further
used to compute the final temperature at the end of the step. This
allows to efficiently reduce noise in the temperature estimation for
optically thin cases. Step 1 allows for a fast convergence of the
temperature but is rather time-consuming when used to derive SEDs,
especially in the low-energy, long wavelength regime. To speed up
convergence, a different scheme (step) is used to compute SEDs.

\item {\bf Step 2} computes the SED  from the 
final temperature distribution calculated in step 1.  In step 2, the number
of photon packets $N_{\gamma
step2}$ is held constant at all wavelengths.
Step 2 therefore maintains a comparable noise level in each wavelength
bins and significantly reduces convergence time by limiting the CPU
time spent in high luminosity bins and focusing on low luminosity
bins. In this step, photon packets are always scattered, no absorption
is allowed. However, at each interaction, the Stokes vector is
weighted by the probability of scattering, $p_\mathrm{sca}$, to take
into account the energy that would have been removed by
absorption. This allows all the photons to exit the disk and to
contribute to the SED, although with a reduced weight.
Packets at a given wavelength $\lambda$ are randomly
emitted from the star and the disk, by cells $i$, with the respective
probabilities~:
\be 
p_* = \frac{L_*(\lambda)}{L_*(\lambda)~+~\sum_i~w_i(\lambda)\,L_i(\lambda)}~,
\ee
\be
{\rm and}~ p_i = \frac{w_i(\lambda)\,L_i(\lambda)}{L_*(\lambda)~+~\sum_i~w_i(\lambda)\,L_i(\lambda)}~.
\ee
where the luminosities are defined by~:
\be
L_*(\lambda) = 4 \pi^2R_*^2\,B_\lambda(T_*)~,
\ee
\be
{\rm and}~ L_i(\lambda) = 4 \pi \, m_i\, \kappa_i^{\mathrm{abs}}(\lambda) \,B_\lambda(T_i)~.
\ee
In order to avoid generating photon packets so deep in the (very
optically thick) disk that
none of them would escape with an appreciable energy left,  a ``dark
zone'' is defined for each wavelength. This dark zone includes all
cells from which the optical depth is at least $\tau_\lambda = 30$, in any
direction, to get out of the computationnal grid\footnote{The value
  $\tau_\lambda = 30$ was found to be a good compromise between
  result accuracy and computational time.}. No
photon packet is emitted from this zone, the emission probability from cells
inside the dark zone being set to zero by fixing a weight
$w_i(\lambda)=0$. A weight $w_i(\lambda) = 1$ is given to cells above the dark zone
limit. Furthermore, packets that would enter
the dark zone during their random walk are killed, because they have
no chance to exit the model volume with significant energy. 
The luminosity of the $N_{\gamma
step2}$ packets emitted at a given wavelength $\lambda$ is determined
by the total energy that the star and the disk radiates at this
wavelength~:
\be
\epsilon_\lambda = \frac{L_*(\lambda)~+~\sum_i~w_i(\lambda)\,L_i(\lambda)}
{N_{\gamma step2}}~.
\ee
\end{itemize}

\subsubsection{Intensity and polarisation maps}

A scheme similar to step 2 used above to calculate SEDs is deployed
in MCFOST to calculate synthetic maps. Maps are calculated, one
wavelength at a time. Scattering is treated explicitly for all
wavelengths but again, absorption is replaced by the appropriate
weighting of the intensity to avoid losses of photon packets.  Images
are produced, in all four Stokes parameters, by classifying photon
packets that exit the calculation grid into inclination bins,
themselves divided into image pixels. The same process applies to
scattered light images in the optical and thermal emission maps in the
far infrared and radio range. The relative contribution of the photospheric and dust thermal
emission is dependent on the wavelength.  It is for instance critical
 in the
  mid-infrared and beyond where it is the (scattered)
thermal emission from the inner disk that starts to dominate over the
photospheric contribution. The correct fraction is calculated
explicitly through the temperature and density of each disk cell and
each contribution can be mapped separately for
comparison. Interferometric visibilities
and phases can be obtained by Fourier transform of images.

\begin{figure}[t]
\centering
\includegraphics[width=8cm,angle=270]{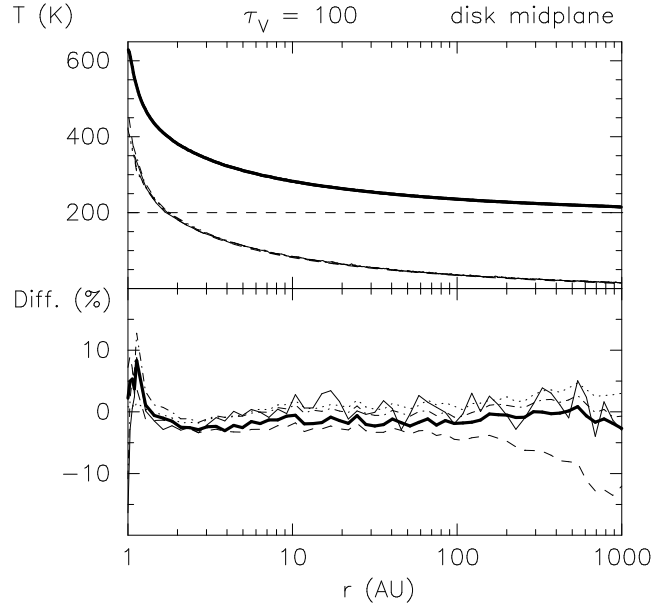}
\caption{Plots of the radial temperature (upper panel) and the
difference (lower panel) for the most optically thick case,
$\tau_\mathrm{V} = 100$, arbitrarily using RADICAL as the
reference.
In both panels, the results of MCFOST are represented by the thick
solid line. Thin solid lines are the results from MC3D, dot-dashed
lines from MCTRANSF, dotted lines from RADMC and dashed lines from
STEINRAY. In the upper panel, because all curves are very similar,
MCFOST has been shifted by 200K for clarity.
\label{fig:Temp_radiale}}
\end{figure}

%======================================================================
\section{\label{sec:test} Tests and validation of MCFOST}

Numerous tests have been performed to check the efficiency, stability,
and accuracy of MCFOST. In this section, we present a comparison of
MCFOST with the benchmark results published by \cite{Pascucci04}
(hereafter P04).

\begin{figure}[t]
\includegraphics[angle=270,width=8.5cm]{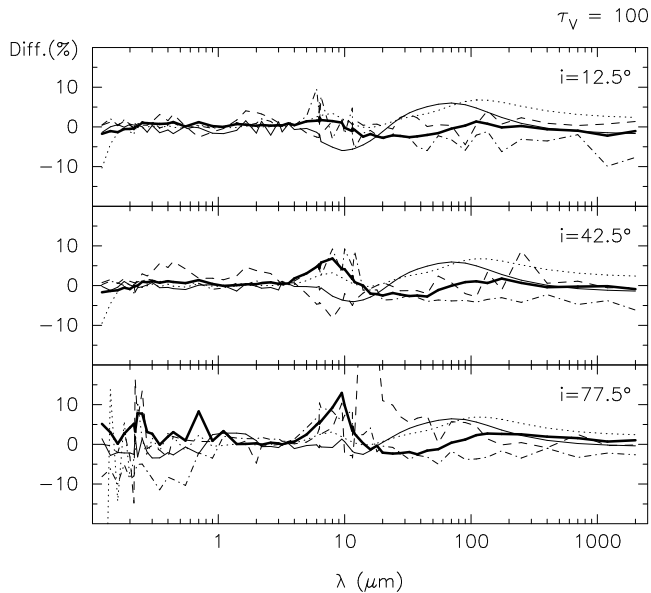}
\caption{\label{fig:diff} Plots of the difference between the model
SEDs for the most optically thick case, $\tau_\mathrm{V} = 100$,
arbitrarily using RADICAL as the reference code. The linestyle-to-code
pairing is the same as in Fig.~\ref{fig:Temp_radiale}.
Three different inclination angles are considered: $i=12.5^\circ$
(upper panel), $i=42.5^\circ$ (middle panel), and $i=77.5^\circ$
(lower panel). 
}
\end{figure}

\subsection{\label{sec:bench} Benchmark definition}
 The geometry tested involves a central point-like source radiating as
a T=$5\,800$K blackbody with a bolometric luminosity $L = L_\odot$
encircled by a disk of well defined geometry and dust content. The
disk extends from 1 AU to 1\,000 AU. It includes spherical dust grains
made of astronomical silicate. Grains have a single size radius of
$0.12\ \mu$m and a density of $3.6\ \mathrm{g.cm}^{-3}$. The optical
data are taken from \cite{Draine84}. The disk is flared with a
Gaussian vertical profile. Radial power-law distributions are used for
the surface density and the scale height (see P04 for more details).
 
To compare with P04, the scattering is also forced to be isotropic and
polarisation is not calculated, i.e, all the scattering information is
contained in the scattering cross section alone, $Q_{sca}$, acting on
the I Stokes parameter only.
 
The number of grid cells is set to $N_r = 50$ and $N_z = 20$ in the
radial and vertical directions, respectively. We used $10^7$ photon
packets to calculate the temperature distribution (step 1) and
$2\,10^6$ photon packets per wavelength to generate the SED (step
2). The total run-time is $20$ minutes for the most optically thick
case on a bi-processor (Intel Xeon) computer running at a clock rate of
2.4 GHz. The run-time memory space needed is $10$ MBytes.

\subsection{Results}

The results calculated with MCFOST for all the test cases presented in
P04 are available at
{\url{http://www-laog.obs.ujf-grenoble.fr/~cpinte/mcfost/}}. Only the
most optically thick case ($\tau_\mathrm{V}=100$, see P04), i.e., the
most difficult one, is presented in Figs.~\ref{fig:Temp_radiale} and
\ref{fig:diff}.

The temperature profiles in the disk and shape of the emerging SEDs
calculated with MCFOST are in excellent agreement with the calculation
presented in P04. Differences in the radial temperatures computed by
MCFOST do not exceed 4\% with respect to the other 5 codes in the
optically thin case. For the most optically thick case, differences
are generally of the order of 2-5\% (see Fig.~\ref{fig:Temp_radiale}).
The results from MCFOST are always within the range of those produced
by other codes. 

Regarding SEDs, MCFOST produces results comparable to all other codes
to better than 10\% for low optical thickness. Deviations as large as
15\% are however observed in the most defavorable case, i.e., high
tilt and optical thickness (either at very short wavelength or in the
10 $\mu$m silicate feature), 
but again the results lie in the range of the values produced by
other codes (see Fig.~\ref{fig:diff}).

Various tests were further performed to show the independence of the
MCFOST results with respect to grid sampling, inclination
sampling, and position of the vertical density cut-off in the disk.
Results are not noticeably affected. A complete description of all
numerical tests is presented in \cite{Greta05}.

\subsection{Energy conservation at high optical depth}
As discussed in section \ref{sec:sampling}, SEDs can be constructed from packets escaping the system in both steps
of the computation (the SED from step 2 being
better converged by construction). As photon packets are generated
independently through different processes in step 1 and 2, our adaptive sampling of the
radiation field method allows an \emph{a posteriori} check of
the accuracy of calculations. 

By integrating over all inclination
angles and wavelengths, 
the emerging flux must correspond in both cases to the star luminosity
(in case of passive heating). This is true by construction for the SED
calculated in step 1. 

Assuming a standard flared disk geometry and ISM like grains, and
varying (via the disk mass) the V band equatorial optical depth from $0.1$ to $10^{9}$, we
find that the luminosities calculated through the 2 methods always
agree to within better than $1\%$.

\section{Example of applications of MCFOST}

In this section we present calculations of the SEDs and infrared
photometric colours of young objects surrounded by circumstellar
disks. The colours calculated by MCFOST are compared with photometric
observations obtained with the {\sl Spitzer Space
Telescope}. Calculated infrared fluxes as a function of disk mass are
also compared with {\sl Spitzer}'s detection limits, as obtained for
example during legacy surveys like {\sl c2d} \citep{Evans03}.  The goals are (i) to explore
whether a disk with ``representative'' geometrical parameters can
match the observed infrared colours of solar-like young stars and (ii)
also to estimate the range of disk masses detectable by {\sl Spitzer}
in such programmes.

\subsection{Disk geometry and central object properties}

\begin{table}[t]
  \centering
  \caption{Star parameters}
  \begin{tabular}{l|ccc}
    Central object & Mass ($M_\odot$) & Radius ($R_\odot$) &
    $T_\mathrm{eff}$\\
    \hline
    T Tauri  star & 1.0 & 1.94 & 4\,420 K\\
    Low mass star  & 0.2 & 1.28 & 3\,365 K\\
    Brown dwarf & 0.05  & 0.57 & 2\,846 K\\
    \hline
  \end{tabular}
  \label{tab:parameters}
\end{table}

\begin{figure*}[t]
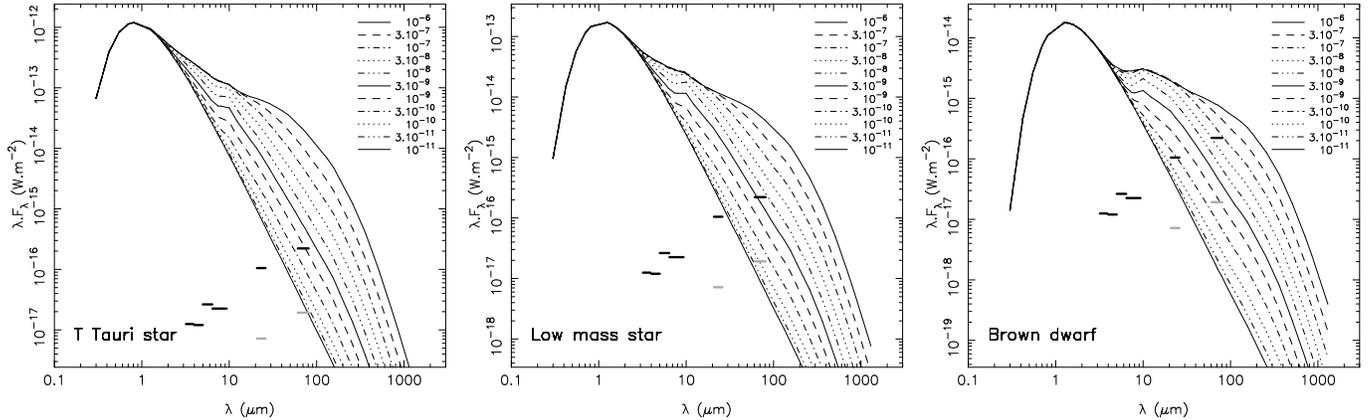

  \includegraphics[angle=270,width=0.325\textwidth]{sed_tt.eps}
  \hfill
  \includegraphics[angle=270,width=0.325\textwidth]{sed_vlms.eps}
  \hfill
  \includegraphics[angle=270,width=0.325\textwidth]{sed_bd.eps}
\caption{\label{fig:SED_Spitzer} SEDs for different dust disk masses
  and central objects, for a pole-on disk. The left panel presents the
  results for the T Tauri star and $R_\mathrm{in}  = 0.1$  AU, the central panel for the very low
  mass star and the right panel for the brown dwarf (defined in table \ref{tab:parameters}). The line type
  corresponds to the disk dust mass (in solar masses). The horizontal
  bars indicate {\sl Spitzer}'s detection limits in the four IRAC
  bands and  the $24\ \mu$m  and  $71\ \mu$m MIPS bands
  for the molecular cloud mapping programme of the {\sl Cores to
  Disks} legacy survey. Additional MIPS limits (light gray bars~:
  $170$ mJy at $24\ \mu$m and $1\,000$ mJy at $71\ \mu$m) for deeper
  pointed observations of the programme are also plotted at $24\ \mu$m
  and $71\ \mu$m.}
\end{figure*}

We assume a disk geometry that is representative of the geometries
derived from images of several disk sources (HK Tau B~: 
\citealp{Stapelfeld98} ; HV Tau C~: \citealp{Stapelfeldt03} ; IRAS 04158+2805~:
\citealp{Glauser06} ; TW Hya~: \citealp{Krist00} ; LkHa 263 C~:
\citealp{Chauvin02}). The disk geometry is kept constant through the
various model runs but the disk mass is varied. Only disks with \emph{dust
masses} smaller than $10^{-6} M_\odot$ are considered in order to focus
on the transition between optically thick and optically thin disks,
{\sl i.e.}, from class II to class III young stellar objects.

The disk is defined by a flared geometry with a vertical Gaussian
profile $\rho(r,z) = \rho_0(r)\,\exp(-z^2/2\,h(r)^2)$.  Power-law
distributions describe the surface density,
$\Sigma(r)=\Sigma_0\,(r/r_0)^{\alpha}$, and the scale height, $ h(r) =
h_0\,(r/r_0)^{\beta}$, where $r$ is the radial coordinate in the
equatorial plane, $h_0 = 10$ AU is the scale height at the reference
radius $r_0 = 100$ AU, $\beta=1.125$, and $\alpha=-1$. The
outer radius is set to $R_\mathrm{out} = 300$ AU.  Two values of
the disk inner radius are considered: $0.1$~AU and $1$~AU.
A grid of models is calculated from dust disk masses ranging from
$10^{-13}$ to $10^{-6} M_\odot$. Calculations are presented with
masses sampled logarithmically, by steps of $\sqrt{10}$.

A grain size distribution $\mathrm{d}n(a) \propto a^{-3.7}
\,\mathrm{d}a$ of spherical particles ranging from $a_\mathrm{min} =
0.03\ \mu$m to $a_\mathrm{max} = 1\ $mm, and optical constants from
\cite{Mathis89}, namely their model A, are used.

 The central stars are assumed to emit like blackbodies. 
Three
different central light sources are considered. They are chosen to
represent a typical T Tauri star, a low mass star, and a brown dwarf
(see Tab.~\ref{tab:parameters} for parameters). The radii and
effective temperatures of the T Tauri and low-mass stars are obtained
from the models of \cite{Siess00}, assuming an age of $2$ Myrs and a
metallicity $Z = 0.1$. For the brown dwarf, a mass of $0.05$
M$_\odot$ is selected. The photospheric properties
are taken from \cite{Chabrier00} and from \cite{Baraffe02} dusty
models. These parameters are summarized in Tab.~\ref{tab:parameters}. 

All
models are calculated for a distance d=140 pc, typical of nearby star
forming regions. 

\subsection{Detectability of disk infrared excesses with {\sl Spitzer}}

In Fig.~\ref{fig:SED_Spitzer}, SEDs are presented as a function of
disk dust mass for the three objects to evaluate the minimum
detectable disk mass as a function of the luminosity of the central
star. For comparison, a set of detection limits reached by {\sl Spitzer} for
large mapping programmes is overplotted. Here we take the values
quoted by the {\sl Spitzer} legacy survey team {\sl Cores to Disks} for
their molecular cloud mapping programme \citep{Evans03} \footnote{See the {\sl c2d} website for
more information \url{http://peggysue.as.utexas.edu/SIRTF/}}.

At all 4 IRAC bands, the photospheric emission is always above the
detection limit, for all types of objects. At $24\ \mu$m, the T
Tauri and low-mass photospheres are detected. However, the photosphere
of the brown dwarf is $\approx 3$ times lower than the detection
limit. This is of course dependent on the exact detection limit chosen
and on the actual mass of the brown dwarf.

For the case considered here, i.e., detection limits for large mapping
programmes and substellar object of 0.05M$_\odot$, a flux detection at
$24\ \mu$m readily implies the presence of circumstellar material. A
more careful analysis is required to confirm a disk excess in the case
of more massive objects. At $71\ \mu$m, the photosphere is detected
for none of the objects and the presence of circumstellar dust can be
directly inferred from any detection.

\begin{figure*}[th]
  \includegraphics[width=8cm]{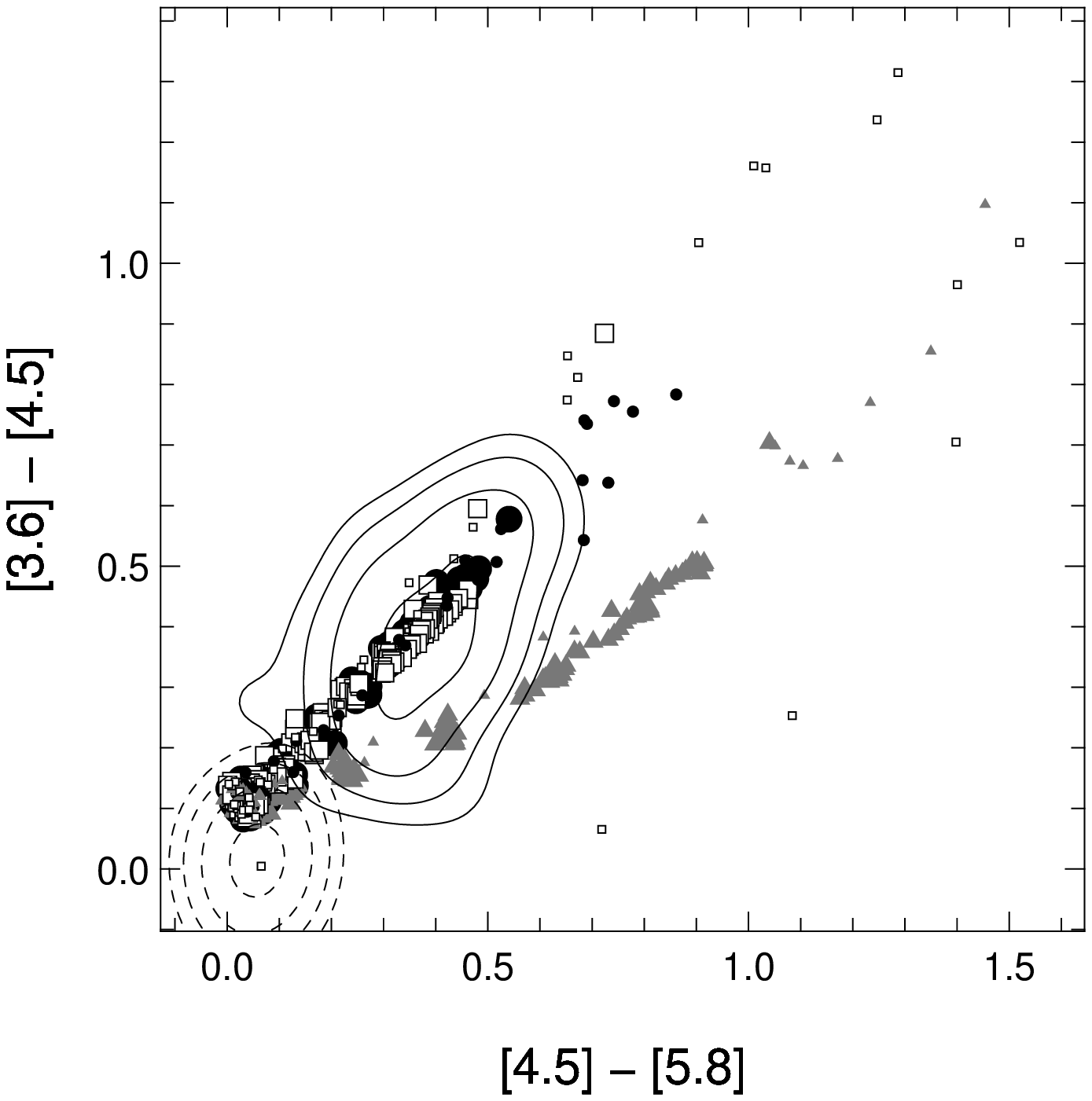} \hfill
  \includegraphics[width=8cm]{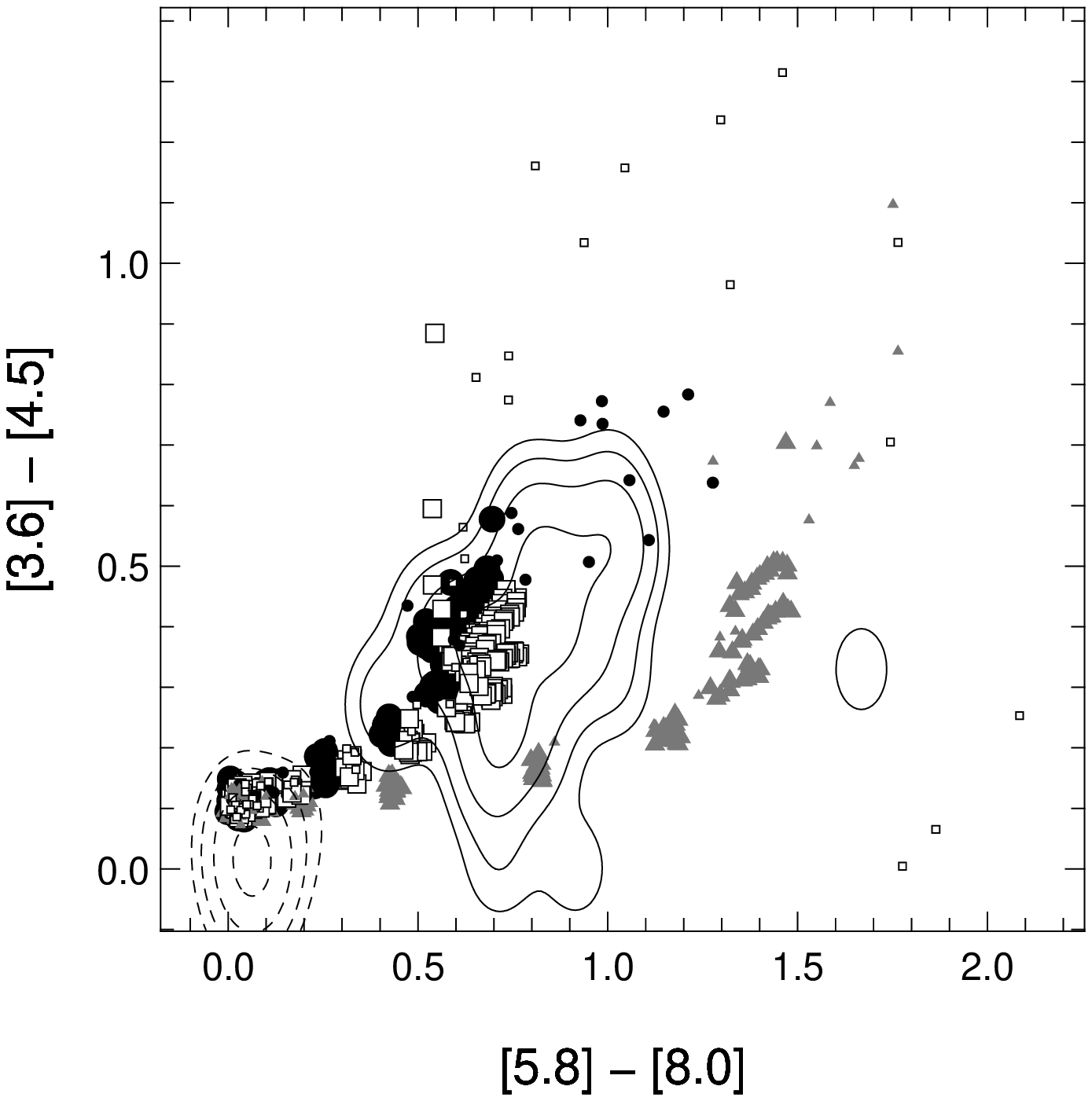} \caption{IRAC colour-colour diagrams. Black filled
  circles corresponds to models with $R_\mathrm{in} = 0.1$ AU and
  large grains ($a_\mathrm{max} = 1$ mm), gray triangles to models
  with $R_\mathrm{in} = 1.0$ AU and large grains, white squares to
  models with $R_\mathrm{in} = 0.1$ AU, with interstellar dust
  ($a_\mathrm{max} = 1\ \mu$m). Large symbols show inclinations
  smaller than $80^\circ$ and small symbols $i > 80^\circ$. Class II
  and Class III data points from \cite{Hartmann05} were converted to
  surface densities using a Gaussian convolution ($\sigma = 0.07$
  mag). The solid and dashed contours define the location of class II
  and class III objects; respectively. The levels are [0.1, 0.2, 0.4,
  0.8] of the maximum of density.}  \label{fig:IRAC_colours}
\end{figure*}

Reverting the argument, for an observing programme with such detection
limits, {\sl Spitzer} is expected to detect, at $71~\mu$m, dust disk masses as low as
$3\,10^{-9}$ M$_\odot$ around T Tauri stars, with $R_\mathrm{in}=0.1$ AU (Fig. \ref{fig:SED_Spitzer}, left panel). For low mass stars, disk
masses greater than $3\,10^{-8}$ M$_\odot$ will be detected (Fig. \ref{fig:SED_Spitzer}, central panel), and for a
brown dwarf of 0.05M$\odot$, disk masses larger than $3\,10^{-7}$
M$_\odot$ are required for
detection (Fig. \ref{fig:SED_Spitzer}, right panel). All results are given in terms of {\sl dust mass}. A
gas-to-dust ratio has to be assumed to extrapolate to total disk
masses.

\subsection{Spitzer colours}
\cite{Hartmann05} have used {\sl Spitzer}/IRAC to determine the
location of TTS (Class II and III) in colour-colour diagrams. Here we
attempt to reproduce the global location of these sources in the
colour-colour diagrams. The precise location depends on the model
geometry.  Similar studies of IRAC colours have been presented
elsewhere \citep{D'Alessio06}. Here, we present the colours of a few
parametric models of passive disks to explore the range of parameters
that reproduce IRAC colours. We also compare the same model with MIPS
colours recently published \citep{Padgett06}.

\begin{figure}[th]
  \includegraphics[width=8cm]{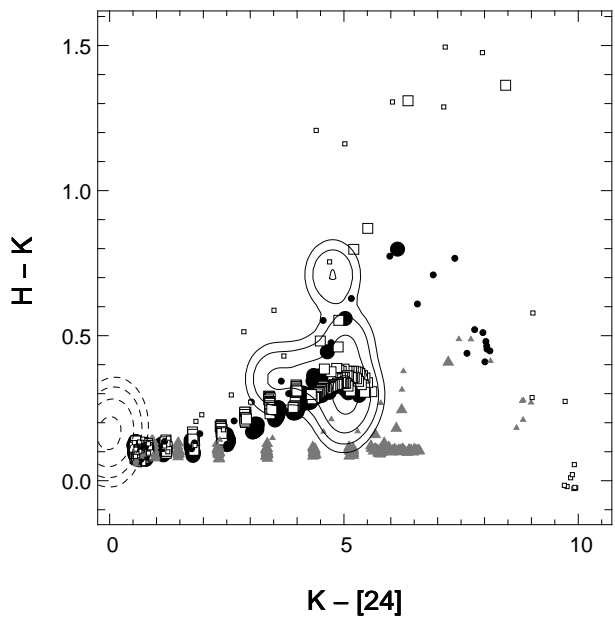}
 \caption{MIPS
  colour-colour diagram. The symbols are the same as in
  Fig.~\ref{fig:IRAC_colours}. The solid and dashed contours define the
  location of class II and class III from \cite{Padgett06},
  respectively. The Gaussian convolution was performed with $\sigma =
  0.07$ mag along the H - K axis and $\sigma = 0.35$ mag along the
  K -[24] axis.
  \label{fig:MIPS_colours}}
\end{figure}

Fig. \ref{fig:IRAC_colours} and \ref{fig:MIPS_colours} show the modelled
IRAC and MIPS colours for the previously defined T Tauri ($1 M_\odot$)
models.  Models for central objects with lower masses (not displayed
for clarity) presents similar colours and, as a consequence, the
comparison of our results for a unique central object mass with the
distribution of masses of \cite{Hartmann05} and \cite{Padgett06}
sources is relevant.
Additional models with interstellar dust, where
$a_\mathrm{max} = 1\ \mu$m and larger inner radius $R_\mathrm{in}=1$
AU are also presented. Three data sets ( $R_\mathrm{in}=0.1$ AU and
$a_\mathrm{max} = 1$\,mm ; $R_\mathrm{in}=1$ AU and $a_\mathrm{max} =
1$\,mm ; $R_\mathrm{in}=0.1$ AU and $a_\mathrm{max} = 1\ \mu$m) are
then displayed, each one calculated for $15$ different disk masses and
for $21$ different inclinations, linearly sampled from pole-on to
edge-on.  Contour plots showing the observed colours of Taurus class II
and class III sources (from \citealp{Hartmann05} in
Fig.~\ref{fig:IRAC_colours} and \citealp{Padgett06} in Fig.~\ref{fig:MIPS_colours}) are surimposed.

Broadly speaking, models sample the two regions of the colour-colour
diagram where sources are found~: low disk mass models corresponding
to class III objects and higher disk mass models to class II. 

The slight offsets seen between the lowest mass models and the class
III region is due to the colours of the blackbody spectrum we used.
Using a more realistic spectrum would produce, for instance, a [3.6] -
[4.5] colour index of $-0.04$ mag instead of $0.1$ mag for the blackbody spectrum, in better agreement with the observations.

For increasing disk masses, all IRAC and MIPS colour indices increase
toward the class II region. The models also populate the intermediate
region where no object are observed. These results can be used to
confirm that the transition between class II (disk dust mass $>
3\,10^{-9}$ M$_\odot$) and class III (disk dust mass $< 10^{-10}$)
M$_\odot$ must be fast enough so that no or very few objects are
detected in samples of nearly 50 sources \citep{Hartmann05}.

A clear difference is seen between models depending on the disk inner
radius~: models with $R_\mathrm{in}=1$ AU produce redder [4.5] -
[5.8] and [5.8] - [8.0] colours.  In particular, with $R_\mathrm{in}=1$
AU the [5.8] - [8.0] colour index becomes larger than the observed
value by about $0.5$ mag for large masses, suggesting that
$R_\mathrm{in}=1$ AU is an upper limit for most class II disks in
Taurus. This suggestion is supported by the H - K / K - [24]
diagram, where no excess is seen in H - K for models with $R_\mathrm{in}=1$
AU, well below the average H - K $=0.3$ seen in observations
(see gray triangles in Fig.~\ref{fig:IRAC_colours} and \ref{fig:MIPS_colours}).
  Such a behaviour is explained by the much higher
temperature reached in models with $R_\mathrm{in}=0.1$ AU, for which
disk emission peaks between $2$ and $3\, \mu$m, whereas it peaks
around $10\, \mu$m in models with $R_\mathrm{in}=1.0$ AU.

Models close to edge-on ($i > 80^\circ$) tend to produce redder colour
at large masses than their less tilted counterparts. Because the
central star becomes more attenuated by its disk, the contribution
from the disk emission progressively dominates, leading to rising
mid-infrared SEDs. The resulting colours are reminiscent of Class I
objects, and highlight the necessity to determine the geometry and
inclination, in order to correctly assess the nature of a class I
source, {\sl i.e.}, the presence or not of a massive envelope
\citep{Kenyon95,White04}

No significant influence of the grains size is detectable for models
where the star is seen directly (with $i < 80^\circ$).  At high
inclinations, on the contrary, models with interstellar dust produce
redder colour indices than models with large grains, for which
absorption tends to be gray.

For completely edge-on models, when no direct light from the star
reaches the observer, short wavelength colours~: [3.6] - [4.5] and H
- K strongly depend on scattering properties.  Models with
interstellar dust present almost no excess at short wavelength, most
of the light being scattered starlight. At longer wavelength, direct
emission from the disk is seen, producing red indices. For models with
larger grains, the albedo remains non negligible in the infrared, and
the contribution of thermal emission from the central (and hidden)
parts of the disk scattered by the outer parts towards the observer is
non negligible and adds to the stellar scattered light, resulting in
redder [3.6] - [4.5] and H - K indices than with interstellar
dust.

%======================================================================
\section{Summary}

We have presented a new continuum 3D radiative transfer code, MCFOST. The efficiency and reliability of MCFOST was tested considering the benchmark configuration defined by P04.
 MCFOST was shown to calculate temperature
distributions and SEDs that are in excellent agreement with previous
results from other codes. 

Sets of models for young solar-like stars and low-stars are presented
and compared to the {\sl Spitzer}'s detection limits for the
programme of the {\sl Cores to Disks} legacy survey.  Minimal disks
masses of $\approx 10^{-9}$ M$_\odot$ for a T Tauri star and $\approx
10^{-7}$ M$_\odot$ for a brown dwarf are needed for the disk to be
detectable by {\sl Spitzer} in the mapping mode.

IRAC and MIPS colours of Taurus Class II and III objects are compared
to passively heated models with a ``representative'' disk
geometry. The average location of the two classes of objects are well
reproduced, as well as the extreme colours of some of the objects, that
may correspond to highly tilted disks.  $R_\mathrm{in} =1$ AU is found
to be maximum inner radius required to account for the observed colours
of Class II objects. Further modelling of individual sources,
combining optical, infrared and millimeter photometry with images
and/or IRS spectroscopy, will be needed to better understand
individual disks, their
dust properties and their evolution.

%======================================================================
\begin{acknowledgements}

Computations presented in this paper were performed at the Service
Commun de Calcul Intensif de l'Observatoire de Grenoble (SCCI). We thank the {\sl Programme National de Physique Stellaire}
(PNPS) and {\sl l'Action Sp\'ecifique en Simulations Num\'eriques pour
l'Astronomie} (ASSNA) of CNRS/INSU, France, for supporting part of
this research. Finally, we wish to thank the referee, C.P. Dullemond, for
his comments which have helped to improve the manuscript.

\end{acknowledgements}

%======================================================================
\bibliographystyle{aa}
\bibliography{biblio}
\end{document}